\documentclass[sigconf]{acmart}

\usepackage{booktabs}
\usepackage{siunitx}

\AtBeginDocument{%
  }

\setcopyright{acmlicensed}
\copyrightyear{2018}
\acmYear{2018}
\acmDOI{XXXXXXX.XXXXXXX}
\acmConference[Conference acronym 'XX]{Make sure to enter the correct
  conference title from your rights confirmation email}{June 03--05,
  2018}{Woodstock, NY}
\acmISBN{978-1-4503-XXXX-X/2018/06}

\begin{document}

\title{Learning from the Storm: A Multivariate Machine Learning Approach to Predicting Hurricane-Induced Economic Losses}

\author{Bolin Shen}
\email{blshen@fsu.edu}
\affiliation{%
  \institution{Florida State Univeristy}
  \city{Tallahassee}
  \country{USA}
}

\author{Eren Erman Ozguven}
\email{eozguven@eng.famu.fsu.edu}
\affiliation{%
  \institution{FAMU-FSU College of Engineering}
  \city{Tallahassee}
  \country{USA}
}

\author{Yue Zhao}
\email{yzhao010@usc.edu}
\affiliation{%
  \institution{University of Southern California}
  \city{Los Angeles}
  \country{USA}
}

\author{Guang Wang}
\email{guang.wang@fsu.edu}
\affiliation{%
  \institution{Florida State Univeristy}
  \city{Tallahassee}
  \country{USA}
}

\author{Yiqun Xie}
\email{xie@umd.edu}
\affiliation{%
  \institution{University of Maryland}
  \city{College Park}
  \country{USA}
}

\author{Yushun Dong}
\email{yd24f@fsu.edu}
\affiliation{%
  \institution{Florida State Univeristy}
  \city{Tallahassee}
  \country{USA}
}

\renewcommand{\shortauthors}{Trovato et al.}

\begin{abstract}
  Florida is particularly vulnerable to hurricanes, which frequently cause substantial economic losses. While prior studies have explored specific contributors to hurricane-induced damage, few have developed a unified framework capable of integrating a broader range of influencing factors to comprehensively assess the sources of economic loss. In this study, we propose a comprehensive modeling framework that categorizes contributing factors into three key components: (1) hurricane characteristics, (2) water-related environmental factors, and (3) socioeconomic factors of affected areas. By integrating multi-source data and aggregating all variables at the finer spatial granularity of the ZIP Code Tabulation Area (ZCTA) level, we employ machine learning models to predict economic loss, using insurance claims as indicators of incurred damage. Beyond accurate loss prediction, our approach facilitates a systematic assessment of the relative importance of each component, providing practical guidance for disaster mitigation, risk assessment, and the development of adaptive urban strategies in coastal and storm-exposed areas. Our code is now available at: \url{https://github.com/LabRAI/Hurricane-Induced-Economic-Loss-Prediction}

\end{abstract}



\keywords{Machine Learning, Storm Surge, Economic Losses, Risk Assessment}

\received{20 February 2007}
\received[revised]{12 March 2009}
\received[accepted]{5 June 2009}

\maketitle

\section{Introduction}
Hurricanes are among the most destructive natural disasters~\cite{davidson2001comparing,west1994modeling,eisenman2007disaster}, 
often causing widespread devastation and severe economic losses~\cite{baade2007estimating,deryugina2018economic,petterson2006preliminary}. 
When making landfall, these storms typically strike coastal regions, where concentrated populations, infrastructure, and economic activity amplify their impact~\cite{greening2006hurricane,wang2016review,sallenger2007coastal}. 
Florida, with one of the longest coastlines in the United States and frequent exposure to Atlantic and Gulf hurricanes, is particularly vulnerable~\cite{frazier2010influence,hung2016vulnerability,chao2021exploratory}. 
Between 1943 and 2024, the state experienced over 500 storms, resulting in a cumulative economic loss exceeding \$300 billion. Given the growing risk posed by climate change and coastal development, it is critical to identify the key factors that contribute most significantly to hurricane-induced damage~\cite{emanuel2006statistical,huang2001long,nofal2021methodology}. 
Understanding these factors can directly inform effective disaster mitigation strategies, enhance the accuracy of risk assessment models, and guide more resilient and adaptive urban planning in regions frequently affected by hurricanes~\cite{zhang2009planning,wolshon2005review}.

To better predict economic losses and support effective risk assessment, a growing body of research has explored various modeling approaches. Traditional methods, such as the Analytical Hierarchy Process (AHP) \cite{mahapatra2015coastal,mansour2021geospatial,saravanan2023multi}, rely heavily on expert knowledge and manual weighting of factors, making them both computationally intensive and subject to subjectivity and inconsistency in large-scale applications. In contrast, machine learning (ML) has emerged as a powerful alternative due to its ability to automatically learn complex patterns from large, heterogeneous datasets.
Recent studies have leveraged ML to analyze specific drivers of hurricane-induced losses. For instance, \cite{arachchige2025ai} investigated the role of housing reconstruction costs in predicting economic losses, while \cite{wendler2022modeling} focused primarily on storm-related meteorological features, such as wind speed and central pressure. Additionally, \cite{yang2024tropical} applied ML techniques to assess the economic impacts of flooding, a common secondary hazard associated with hurricanes.
Despite these advancements, there remains a gap in the literature: no existing work has holistically integrated the full spectrum of contributing factors. In particular, few studies consider the combined influence of (i) hurricane characteristics, (ii) water-related environmental factors, and (iii) socioeconomic factors of affected regions. This omission limits our understanding of how these factors interact to shape the severity of hurricane damage.

To address this gap and provide a more comprehensive understanding of the drivers behind hurricane-induced economic losses, we propose a framework that categorizes contributing factors into three main dimensions. First, we characterize the intrinsic properties of each hurricane using meteorological variables, including maximum sustained wind speed, hurricane category, and minimum central pressure. These features capture the storm’s physical intensity and destructive potential.
Second, we consider the hydrological vulnerability of the affected regions by quantifying the number of water-related entities present in the landfall zone. These features serve as proxies for the risk of secondary hazards such as storm surge and inland flooding.
Third, we incorporate human-related vulnerability by extracting structural attributes from the built environment, such as elevation difference, which quantifies how much higher or lower a building sits relative to the base flood elevation and acts as a significant factor affecting storm-induced losses.
We use total replacement cost, which combines the estimated value of buildings and contents reported by insurers, as a quantifiable measure of economic loss. All variables are aggregated at the ZIP Code Tabulation Area (ZCTA) level to support spatial analysis. By applying machine learning models to this integrated dataset, we not only achieve accurate predictions of economic loss but also evaluate the relative importance of each category of contributing factors. These insights provide meaningful guidance for disaster mitigation, risk management, and urban resilience planning in regions exposed to severe storm impacts. Our contributions are as follows:

\begin{itemize}
    \item We propose a unified framework that integrates hurricane characteristics, environmental factors, and socioeconomic factors to comprehensively model economic losses.
    \item We assess various machine learning models and demonstrate their consistently strong performance in predicting hurricane-induced economic losses.
    \item We analyze the relative importance of contributing factors, providing an evidence base for risk assessment and actionable insights for mitigating hurricane-induced losses.
\end{itemize}

\section{Methodology}
This section describes the key variables selected to predict economic loss, the procedures for data processing and feature construction, and the machine learning models used for predictive analysis.

\subsection{Data Preparation}

The variables of interest are organized into three conceptual domains, encompassing hurricane characteristics, water-related environmental factors, and socioeconomic conditions, to capture the diverse drivers of hurricane-induced economic loss.

\subsubsection{Hurricane Characteristics}
We extracted storm-specific features from the IBTrACS dataset to quantify hurricane intensity and structure. Specifically, we included the maximum sustained wind speed, the corresponding hurricane category based on the Saffir-Simpson Hurricane Scale, and the minimum recorded sea-level pressure. These variables reflect the physical force and severity of each hurricane, which are known to strongly influence the extent of resulting damage.

\subsubsection{Water-Related Environmental Factors}
To capture the hydrological characteristics of affected areas, we incorporated features from the Florida Hydrography Dataset (FHD). For each ZCTA, we counted the number of dams, outlets, hydrological stations, and streamgages. These elements serve as proxies for the presence of water-related infrastructure and the region’s susceptibility to secondary disasters such as flooding, which may significantly exacerbate the economic impact of a hurricane.

\subsubsection{Socioeconomic Factors}
We obtained information about the built environment and human exposure from the OpenFEMA dataset. The selected features capture key aspects of structural vulnerability and exposure, including: the average building age, derived from the original year of construction; the average number of building floors, reflecting vertical development intensity; the average elevation difference, which measures a structure’s vertical distance from the base flood elevation; and the average number of elevated buildings, indicating the prevalence of elevation-based flood mitigation. Together, these attributes provide essential context for understanding socioeconomic variation in hurricane-induced losses.

\subsection{Feature Engineering}
We integrated features from three domains, including hurricane characteristics, water-related environment factors, and socioeconomic factors, at the ZCTA level to construct a unified representation for prediction. Each ZCTA was assigned storm features based on the geographically nearest hurricane, reflecting the irregular and sparse distribution of storm events. For hydrological data, which were available at a finer spatial resolution than the loss records, we retained only ZCTAs present in the economic loss data and filled missing values using information from the nearest ZCTA by geographic distance. Socioeconomic features were already reported at the ZCTA level and were directly aligned with the loss information.
All features were then processed to ensure compatibility with the machine learning models. Numerical variables such as building age, elevation difference, and hydrological counts were standardized to have zero mean and unit variance. Ordinal features like hurricane category were preserved in their original form without scaling. Categorical variables such as occupancy type were transformed using one-hot encoding to capture class distinctions without introducing implicit ordering. After preprocessing, the final feature matrix was used as input for all subsequent predictive modeling.
The target variable for prediction is the total replacement cost, composed of both building replacement cost and contents replacement cost as reported by insurers. Since both components reflect monetary values at the time of loss, we adjusted them using the Florida House Price Index (FLSTHPI) to account for inflation\footnote{Adjusted to a baseline of January 1, 2025, at which point FLSTHPI was 820.29.}. Given the high variance in replacement costs across ZCTA regions, we applied a log transformation to stabilize the distribution.

\subsection{Machine Learning Models}
To model the relationship between the constructed features and economic loss, we employed a range of machine learning algorithms, including Random Forest (RF), Gradient Boosting Machine (GBM), Extreme Gradient Boosting (XGBoost), Neural Networks (NN), and a Stacked Ensemble model that combines the predictions of the aforementioned base learners. 
%
We randomly split the full dataset into a training set (80\%) and a validation set (20\%). To ensure robust evaluation, we performed 5-fold cross-validation repeated five times, reporting the average performance along with the standard deviation. In addition to conventional machine learning models, we evaluated a neural network model and further explored a stacked ensemble approach that integrates the strengths of all individual models. This comprehensive experimental design allows for a thorough comparison and provides a solid foundation for identifying the most effective model for predicting economic losses.

\section{Experiments}
This section addresses three central research questions concerning the effectiveness and interpretability of our proposed predictive framework for hurricane-induced economic loss: RQ1: How effective is our comprehensive feature set in predicting economic loss? RQ2: Which feature domain contributes most significantly to prediction accuracy? RQ3: What actionable insights can be drawn from the model to reduce future economic loss?

\subsection{Experimental Settings}
\subsubsection{Model Parameters}
We implemented four predictive models with fixed hyperparameters selected for stable performance. The Random Forest model used 100 trees with a maximum depth of 10. The XGBoost model employed 100 boosting rounds, a learning rate of 0.1, and a maximum depth of 4. The Neural Network consisted of a single hidden layer with 100 units, trained with early stopping and an L2 regularization strength of 0.01. The Gradient Boosting Machine was configured with 100 estimators, a learning rate of 0.1, and a shallow tree depth of 4 to promote generalization.

\subsubsection{Evaluation Metrics}
Model performance was evaluated using five metrics that collectively reflect both absolute and relative predictive accuracy. The R-squared score measures the proportion of variance explained by the model, with higher values indicating better fit. Mean Absolute Error (MAE) captures the average prediction error magnitude, while Root Mean Squared Error (RMSE) emphasizes larger errors through squared differences. Symmetric Mean Absolute Percentage Error (SMAPE) quantifies relative error as a percentage, and Root Mean Squared Logarithmic Error (RMSLE) is particularly suited for targets with wide dynamic ranges. For all metrics except R-squared, lower values indicate better performance.

\subsection{Economic Loss Prediction}

\begin{table}[htbp]
\centering
\caption{Performance metrics of machine learning models for predicting hurricane-induced economic loss.}
\label{tab:prediction}
\setlength{\tabcolsep}{1.5pt}
\begin{tabular}{lccccc}
\toprule
\textbf{Model} & {$R^2$} & {MAE} & {SMAPE} & {RMSE} & {RMSLE} \\
\midrule
NN              & $0.213 \pm 0.1$ & $2.075 \pm 0.2$ & $11.58 \pm 0.7$ & $2.777 \pm 0.2$ & $0.144 \pm 0.0$ \\
RF              & $0.675 \pm 0.0$ & $1.341 \pm 0.1$ & $7.437 \pm 0.5$ & $1.786 \pm 0.1$ & $0.091 \pm 0.0$ \\
XGBoost         & $0.671 \pm 0.0$ & $1.339 \pm 0.1$ & $7.410 \pm 0.5$ & $1.793 \pm 0.1$ & $0.092 \pm 0.0$ \\
GBM             & $0.677 \pm 0.1$ & $\mathbf{1.336 \pm 0.1}$ & $\mathbf{7.371 \pm 0.6}$ & $1.774 \pm 0.2$ & $0.091 \pm 0.0$ \\
Stacked         & $\mathbf{0.684 \pm 0.1}$ & $1.353 \pm 0.1$ & $7.490 \pm 0.6$ & $\mathbf{1.754 \pm 0.1}$ & $\mathbf{0.090 \pm 0.0}$ \\
\bottomrule
\end{tabular}
\end{table}

To answer RQ1, we employed machine learning models to predict hurricane-induced economic loss, and the results are summarized in Table~\ref{tab:prediction}. As shown, tree-based models such as Random Forest, XGBoost, and Gradient Boosting Machines achieved consistently strong predictive performance. This indicates that the selected feature set effectively captures key determinants of hurricane-related damages, enabling accurate and robust predictions. 
In contrast, the neural network model underperformed with an $R^2$ of only 0.213 and considerably higher error metrics. This may be attributed to the relatively small number of hurricane events in the training data, which limits the neural network’s ability to generalize and learn complex patterns effectively. 
The stacked ensemble model, which integrates predictions from all base models, achieving the highest $R^2$ value and the lowest RMSE and RMSLE. This suggests that model ensembling can effectively leverage complementary strengths across individual models, leading to more accurate and stable predictions. 
In conclusion, machine learning models provide a powerful approach to modeling hurricane-induced economic losses. Among them, the stacked ensemble model stands out as the most reliable and precise, offering a promising tool for disaster impact assessment and mitigation planning.

\subsection{Importance Analysis}
\label{section:importance_analysis}

\begin{figure}[htbp]
    \centering
    \includegraphics[width=0.48\textwidth]{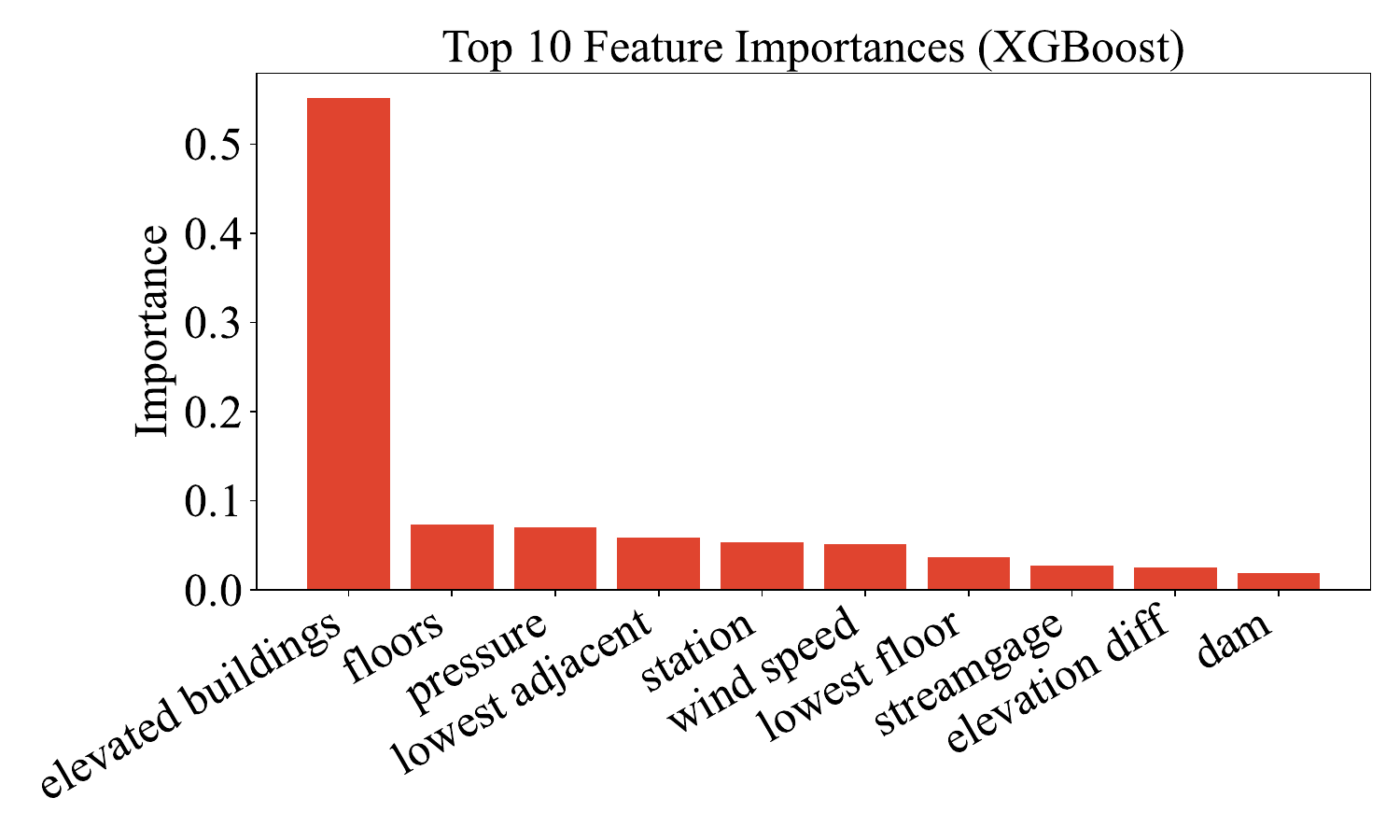}
    \caption{Feature importances from the XGBoost model for hurricane loss prediction.}
    \label{fig:importance}
\end{figure}

To investigate RQ2, which features most significantly contribute to predicting hurricane-induced economic losses, we conducted a feature importance analysis using the XGBoost model. As shown in Fig.~\ref{fig:importance}, the variable representing the number of elevated buildings exhibits the highest importance, indicating that elevation-based structural adaptation is a key determinant of economic damage among human-related factors. Moreover, both the number of floors and the height of the lowest floor also play critical roles in predicting economic losses, highlighting their significance as socioeconomic factors influencing vulnerability and resilience.
Additionally, intrinsic hurricane characteristics such as minimum sea level pressure and maximum sustained wind speed also demonstrate relatively high importance in the prediction task. Regarding local hydrological features, various water-related entities contribute similarly to the model’s performance, indicating that each plays a role in influencing economic losses.
These results suggest that regions with a high number of elevated buildings, more severe hurricane conditions, and greater presence of water-related infrastructure are more likely to incur substantial economic damage. Therefore, targeted hurricane preparedness measures and carefully designed insurance strategies are particularly crucial in these high-risk areas.

\subsection{Disaster Mitigation Insights}

\begin{figure}[htbp]
    \centering
    \includegraphics[width=0.48\textwidth]{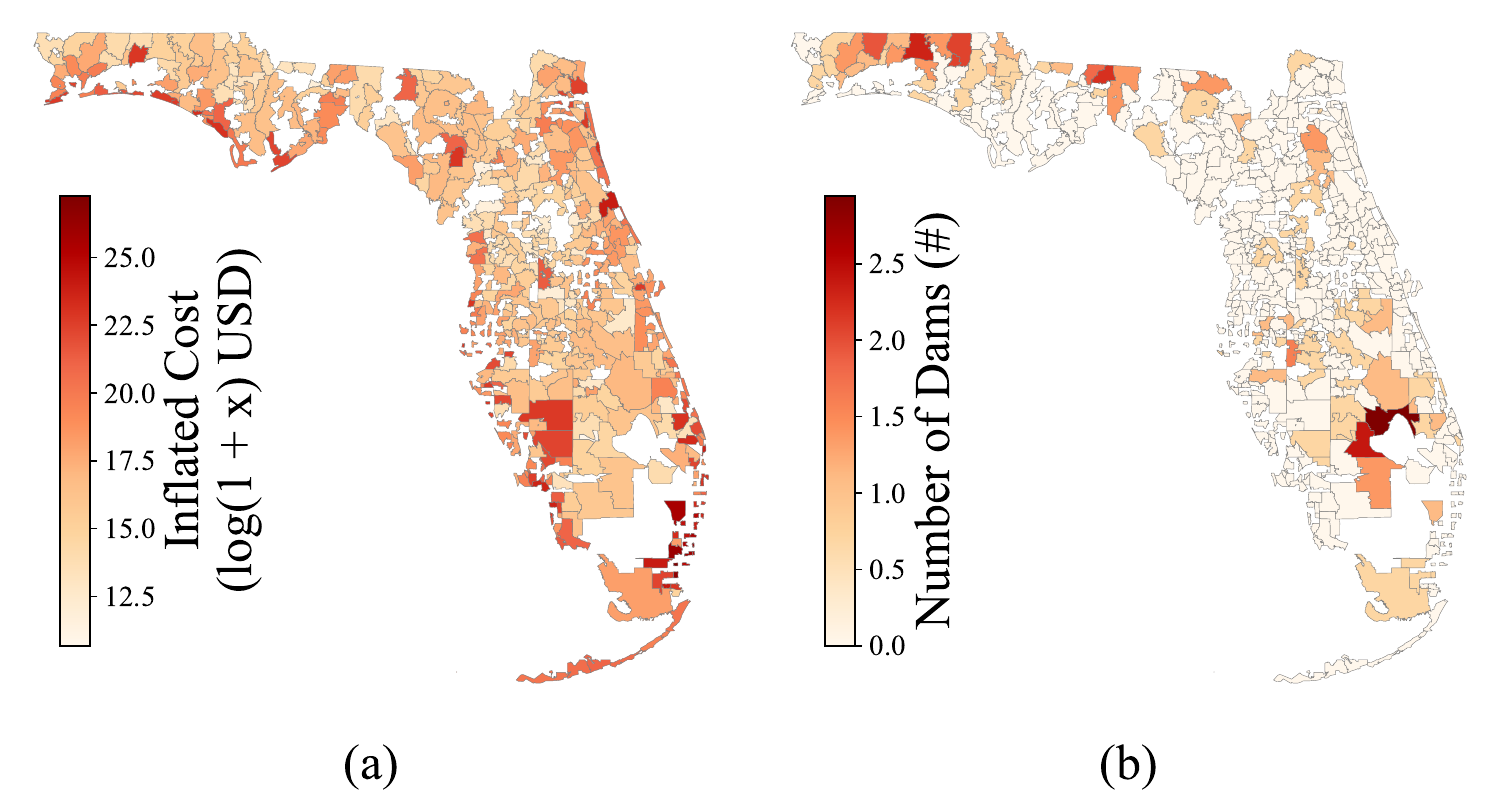}
    \caption{Spatial distribution of selected features at the ZIP Code Tabulation Area (ZCTA) level in Florida. Subfigure (a) shows the inflated total replacement cost. Subfigure (b) presents the number of dams within each ZCTA region.}
    \label{fig:heatmap-2}
\end{figure}

To address Research Question 3, we conducted a detailed analysis at the ZIP Code Tabulation Area (ZCTA) level across the state of Florida. As illustrated in Fig~\ref{fig:heatmap-2}(a), the most significant economic losses are concentrated in coastal areas, particularly along the northwestern and southern coastlines, where the distribution of socioeconomic factors closely aligns with these high-loss regions. Moreover, as shown in Fig~\ref{fig:heatmap-2}(b), water-related environmental factors are primarily concentrated in inland regions, they also serve as important contributors to economic damage, potentially triggering secondary disasters that further exacerbate financial losses. In addition, historical hurricane data indicate that severe storm characteristics are predominantly observed in the northwestern and central parts of Florida, which again corresponds closely with the spatial distribution of economic losses.
Based on these spatial patterns, several urban planning insights can be derived. First, coastal areas in Florida are especially susceptible to hurricane impacts and tend to suffer the most severe economic damage. These areas should therefore be prioritized for the development of coastal defense infrastructure and emergency preparedness strategies. Second, inland regions that contain a high density of water-related infrastructure, such as dams, are at risk of secondary disasters triggered by hurricanes. Targeted measures should be implemented to protect these hydrological facilities. Finally, areas that are historically exposed to high-intensity hurricanes should adopt comprehensive mitigation strategies to reduce their vulnerability. These findings provide valuable guidance for disaster risk reduction, infrastructure planning, and the formulation of region-specific insurance policies.

\section{Conclusion}
In this study, we presented a multivariate machine learning framework for predicting hurricane-induced economic losses at the ZCTA level in Florida. By integrating hurricane characteristics, water-related environmental factors, and socioeconomic attributes, the proposed approach captures the multifaceted drivers of disaster impact.
Through rigorous evaluation of several machine learning models, we demonstrated robust and consistent predictive performance. In addition, the feature importance analysis highlights the relative contributions of different factor domains, offering actionable insights for disaster risk assessment and loss mitigation.
%
Overall, our proposed framework offers reliable loss prediction and actionable insights to support proactive mitigation.


\bibliographystyle{ACM-Reference-Format}
\bibliography{refs}


\end{document}